\documentclass[journal]{IEEEtran}
\usepackage{graphicx, epstopdf}
\usepackage{amssymb, amsmath}
\usepackage{commath, cite}
\usepackage{setspace}
\usepackage{hyperref}
\usepackage{graphicx,cite,amsmath,amssymb,hhline}

\usepackage{xcolor}

\IEEEoverridecommandlockouts

\begin{document}

\title{Simultaneous Information and Energy Flow for \\
IoT Relay Systems with Crowd Harvesting}

\author{Weisi Guo, Sheng~Zhou\textsuperscript{*}, Yunfei~Chen, Siyi~Wang, Xiaoli~Chu, Zhisheng~Niu

\thanks{\textsuperscript{1}Weisi Guo and Yunfei Chen are with the School of Engineering, University of Warwick, UK. \textsuperscript{2}Sheng Zhou (\textsuperscript{*}corresponding author) and Zhisheng Niu are with the Department of Electronic Engineering, Tsinghua University, China. \textsuperscript{3}Siyi Wang is with the Department of Electrical and Electronic Engineering, Xi'an Jiaotong - Liverpool University, China. \textsuperscript{4}Xiaoli Chu is with the Department of Electronic and Electrical Engineering, University of Sheffield, UK.

This work is sponsored in part by the Nature Science Foundation of China under the Grant No.61571265, 61461136004 and 61321061, the Young Scholar Programme of the Jiangsu Science and Technology Programme under the Grant No.BK20150374.

} }

\IEEEoverridecommandlockouts

\maketitle

\begin{abstract}
It is expected that the number of wireless devices will grow rapidly over the next few years due to the growing proliferation of Internet-of-Things (IoT). In order to improve the energy efficiency of information transfer between small devices, we review state-of-the-art research in simultaneous wireless energy and information transfer, especially for relay based IoT systems. In particular, we analyze simultaneous information-and-energy transfer from the source node, and the design of time-switching and power-splitting operation modes, as well as the associated optimization algorithms. We also investigate the potential of crowd energy harvesting from transmission nodes that belong to multiple radio networks. The combination of source and crowd energy harvesting can greatly reduce the use of battery power and increase the availability and reliability for relaying. We provide insight into the fundamental limits of crowd energy harvesting reliability based on a case study using real city data. Furthermore, we examine the optimization of transmissions in crowd harvesting, especially with the use of node collaboration while guaranteeing Quality-of-Service (QoS).
\end{abstract}

\section{Introduction}
Today, wireless services are dominated by packet data transfer over the cellular and Wi-Fi networks. Cellular networks account for the majority of the world's wireless power consumption, with 6 million macro-cells world wide consuming a peak rate of 12 billion Watts. Whilst video demand drives most of the data consumption, machine-2-machine (M2M) data is the fastest growing driver. The rapid growth in the Internet-of-Things (IoT) sector is set to increase the energy consumption of small devices dramatically. Whilst many such small devices are sensor motes with a power consumption that is in the order of a Watt or less, the sheer number of such devices (25 billion) is set to consume more power than the cellular networks worldwide. Therefore, it is important to address the emerging challenge of energy efficiency for connected small devices. In order for small devices to communicate without a tether, wireless relaying has been widely employed in current and emerging wireless systems. For example, M2M-R relaying has been proposed as a suitable heterogeneous architecture for either the ETSI M2M, 3GPP MTC, or 3GPP 802.16p IoT architectures \cite{M2MRelay13}.

Conventionally, relaying operations cost the relaying nodes extra energy and therefore, may prevent battery-operated nodes from taking part in relaying. Thus, RF powered relaying is a promising solution, whereby the relay nodes can harvest energy from either the source node directly \cite{Nasir} or external sources \cite{Mede} (i.e., \emph{crowd harvesting} from external radio transmissions) for sustainable operation, as shown in Fig~\ref{fig-YC1}. In fact, the feasibility of crowd harvesting is supported by the dramatic increase in the density of RF transmitters in cities. For example, the global cellular infrastructure consists of more than 6 million base stations (BSs) and serves more than 7 billion user equipments (UEs), and the number of Wi-Fi access-points (APs) has reached 350 per km$^2$, with many metropolitan areas reaching over 700 per km$^2$\footnote{http://www.smallcellforum.org/press-releases/small-cells-outnumber-traditional-mobile-base-stations}.

This paper addresses several challenges faced by energy harvesting relay systems attempting to optimise throughput. We break down the problem into two domains: i) the source of the RF energy, and ii) optimising the transfer of information subject to different energy harvesting scenarios. In particular, we answer two important research questions: how to optimally schedule information and energy transmission from a common transmitter node, and how much energy can be crowd-harvested from ambient RF signals for relaying?

The organization of the paper is as follows. In Section \ref{one}, we review state-of-the-art research on RF powered relaying systems. In Section \ref{two}, we review the potential of harvesting a crowd of transmission nodes that belong to multiple heterogeneous networks. In Section \ref{three}, we review the optimization of transmissions with crowd harvesting, especially with node collaboration while guaranteeing a certain QoS. Finally, in Section \ref{four}, we discuss open challenges research opportunities in this area.

\section{RF Powered Relaying}
\label{one}

\subsection{SWIPT: Harvesting from the Source Node and Relaying}

If the relay node harvests energy from the source node's RF transmissions, then the relaying becomes a simultaneous wireless information and power transfer (SWIPT) system \cite{Nasir}, as the relay node also receives information from the source node. In this case, there are two main algorithms to implement wireless powered relaying (WPR): \emph{time-switching (TS)} and \emph{power-splitting (PS)}. In TS, the source node transmits energy to the relay node for $\alpha T$ seconds and the remaining $(1-\alpha)T$ seconds are used for information delivery, where $0\le\alpha\le 1$ and $T$ is the total duration of transmission. In PS, a portion $\rho$ of the received power from the source is used for energy harvesting while the rest $1-\rho$ is used for information decoding, where $0\le\rho\le 1$.

\begin{figure}
\begin{center}
\includegraphics[width=1\linewidth]{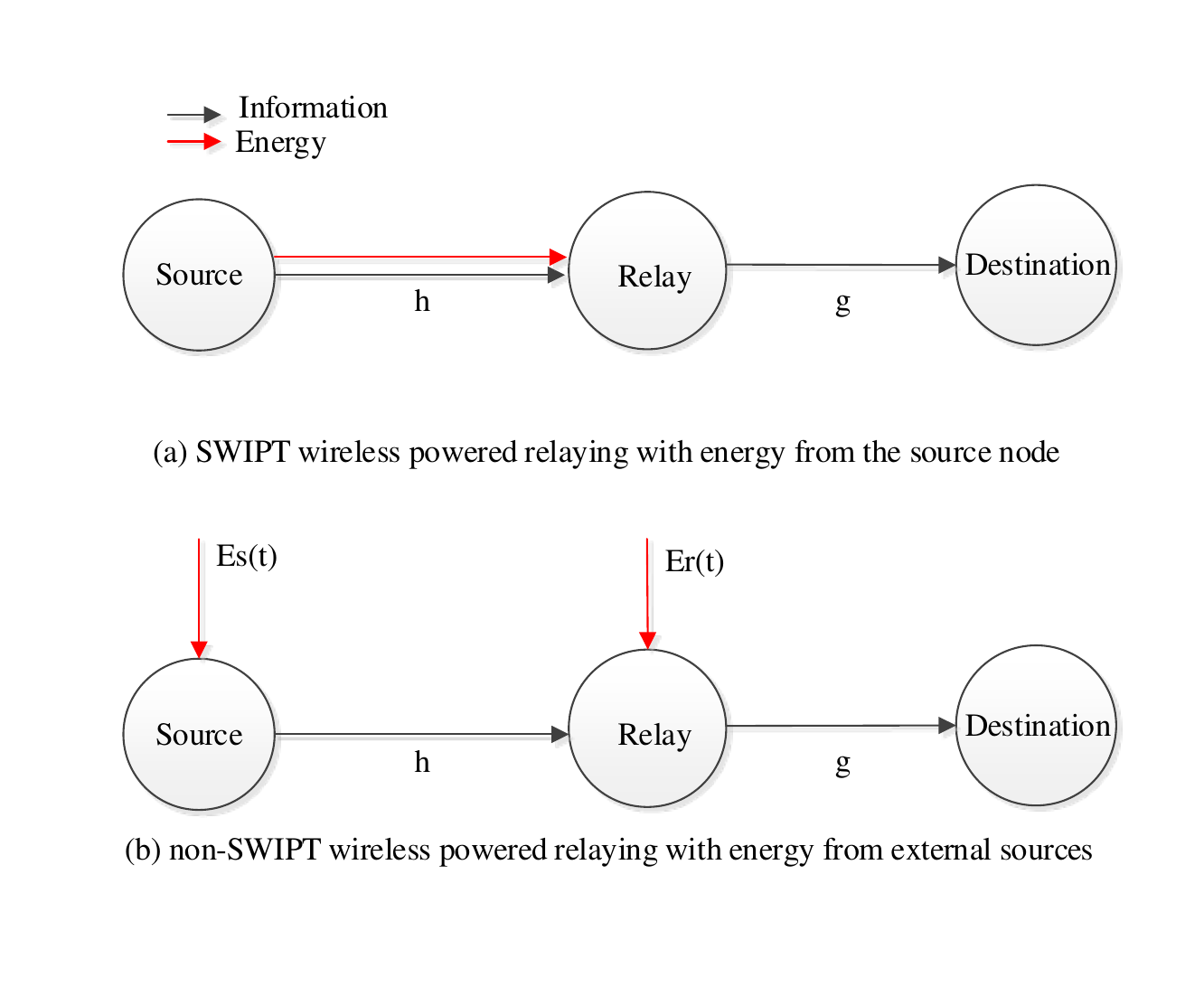}
\caption{Diagram of wireless powered relaying harvesting the source node or harvesting external sources.}
\label{fig-YC1}
\end{center}
\end{figure}

Both TS and PS relaying require a separate energy harvester in addition to the data transceiver at the relay. On the one hand, PS is slightly more complicated in hardware, as it uses a power splitter which is not a trivial hardware component. On the other hand, TS requires a dedicated harvesting time, which adds complexity to synchronization and also reduces the effective throughput of the system. Consequently, under similar conditions, TS often has a smaller throughput than PS. Another issue with PS is that it uses the same signal for energy harvesting and information delivery in the broadcast phase. This could be a problem for amplify-and-forward (AF) relaying, as it uses the harvested energy to forward the rest of the signal directly without any further processing. If the value of $\rho$ is small, a small amount of harvested energy would be used to forward a strong signal; and if the value of $\rho$ is large, a large amount of harvested energy would be used to forward a weak signal, both leading to a weak forwarded signal. Thus, PS often has a smaller transmission range than TS in AF relaying.

\begin{figure}
\begin{center}
\includegraphics[width=1\linewidth]{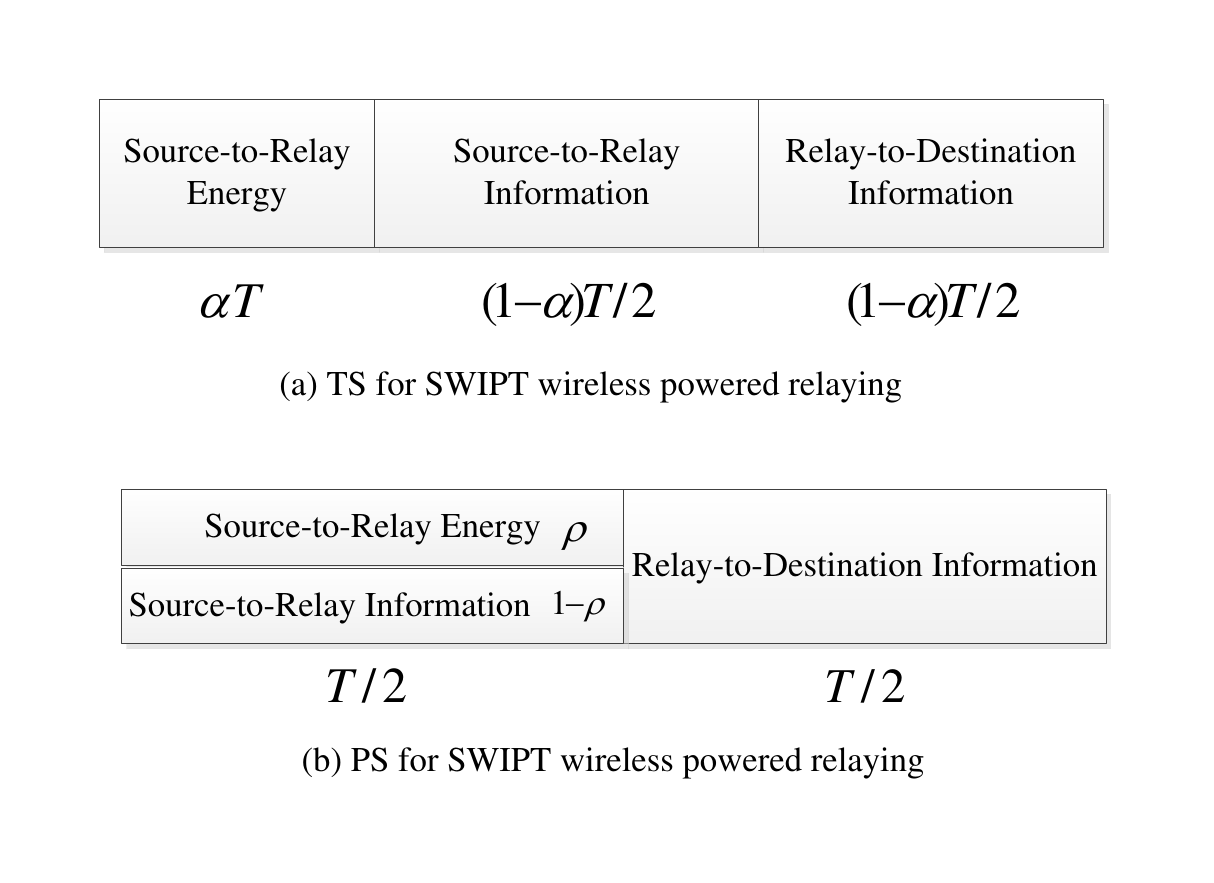}
\caption{TS and PS for SWIPT wireless powered relaying. }
\label{fig-YC2}
\end{center}
\end{figure}

Researches in TS or PS WPR systems have mainly focused on the analysis of the relaying performance and its optimization with respect to $\alpha$ and $\rho$. Fig. \ref{fig-YC2}(a) and Fig. \ref{fig-YC2}(b) show the scheduling of SWIPT relaying. The SWIPT relaying system in \cite{Ion} is different in that the relay node can harvest energy from large-scale network interference or from self-interference in full duplex nodes. Multiple antennas can be employed at either the source node or the relay node to perform beamforming or antenna selection for diversity gain. These are not discussed in detail here.

\subsection{Non-SWIPT: Harvesting from External Sources}

Another type of RF powered relaying systems harvest energy from external sources (e.g., crowd RF transmissions). As the energy and information come from different sources, they do not assume a standard SWIPT structure. However, since the transmission times of the nodes in the same network are normally scheduled by a network-level central controller, the relay can use the transmission times of all other nodes as dedicated harvesting time, and energy is still controlled and correlated with information. Traffic prediction and knowledge of the nodes' transmission schedule will help harvesting node to build up a statistical understanding of the energy arrival intensity and frequency.

To achieve efficient scheduling, one needs to model the energy arrival or the energy profile as a function of time. This model could be a random process, for example, a Markov process that considers energy state transition \cite{Mede}. It could be a probabilistic model, for example, a Bernoulli energy injection model with a probability of $p$ that an energy of $E$ is harvested and a probability of $1-p$ that no energy is harvested, or a simple model with a probability of $1/3$ that no energy is harvested, a probability of $1/3$ that an energy of $E$ is harvested and a probability of $1/3$ that an energy of $2E$ is harvested \cite{Mede}. It could also be a deterministic model where the amount and the arrival time of the energy are known in advance before scheduling \cite{Huang}.

Using these energy models, scheduling can be formulated as an optimization problem that searches for the best transmission time and the best transmission power. Depending on whether the source node and/or the relay node conduct energy harvesting at the $k$-th time slot ($k=1,2,\cdots,K$), the available transmission energy $E_s(k)$ and $E_r(k)$ at the source node and the relay node respectively will be variable. The optimization is bounded by the energy causality, where the transmission power $P_s(k)$ and $P_r(k)$ at the source and relay nodes respectively must be smaller than the available energy from energy harvesting. Moreover, there is an information causality constraint for relaying systems, where the information must be transmitted from the source node before it can be forwarded by the relay node. The available energies at the source and the relay are updated after each transmission. Existing optimization techniques, such as convex optimization, non-convex mixed-integer non-linear program, directional water-filling, and dynamic programming, can be used to find the optimum transmission times and power levels for the source and relay nodes to maximize the sum throughput.

The optimization can be performed for online algorithms that only have causal knowledge of the channel state information (CSI) and energy state information (ESI) as well as for off-line algorithms that assume knowledge of the CSI and energy for all transmissions. For relaying systems, off-line algorithms are very complicated, as they require knowledge of transmissions in at each of the hops. It can be performed by energy-constrained relay nodes that have limited energy as well as energy-unconstrained relay nodes that have unlimited energy. The size of energy storage can be limited or infinite, which may impose an averaging-out effect on the available energy that has been harvested. Even though not practical, the off-line algorithms can provide a performance upper-bound for the energy harvesting relay system, or can be applied when the energy arrival can be predicted to some extent. For online algorithms, the optimal scheduling policy can be obtained via policy iteration with the Markov decision process (MDP) formulation of the problem, especially for nodes with finite energy storage. While the general policy iteration suffers from the curse of dimensionality, i.e., high complexity, heuristic algorithms like threshold-based transmission policies are more practical, and in many cases, their performance can be qualitatively analyzed. The optimization can cover various QoS requirements of the traffic being delivered. For example, it can be performed for delay-constrained case when the relay node must forward the information upon reception, or for non-delay-constrained case when the relay node does not need to forward the information immediately after reception. In the non-delay-constrained case, the binary indicator $d_s(k)=0$ ($d_r(k)=0$) means no information transmission in the $k$-th frame and $d_s(k)=1$ ($d_r(k)=1$) means information transmission in the $k$-th frame at the source (relay) node. Alternatively, one can minimize the total relay transmission time instead of maximizing its throughput.

Furthermore, the SWIPT and non-SWIPT modes can be combined together. For example, a SWIPT mode could be activated when insufficient energy is harvested from the non-SWIPT mode. However, this combination brings challenges too. First, since the relaying system and the ambient systems may not operate on the same frequencies, multiple energy harvesters may be required with increased hardware complexity. Second, since the activity of one mode depends on the other mode, the optimization of $\alpha$ and $\rho$ in SWIPT and the optimization of transmission power and time in non-SWIPT will be technically difficult.

\section{Crowd Harvesting from External Radio Transmissions}
\label{two}
There has been a rapid growth in the density of fixed and mobile wireless devices globally. Whilst the increase in transmitter density will undoubtedly increase the amount of RF energy available in urban environments, it is difficult to quantify the amount of energy reliably available for any given energy harvesting device over some arbitrary time period. The lack of certainty is due to three complexities, namely: i) random nature of RF transmitter locations, ii) randomness in the RF propagation channel, and iii) variations in spectrum utilization due to traffic patterns.

%All of the above factors affect the amount of energy available, which needs to be converted into constant DC voltage for usage. At the moment, low-power devices with RF energy harvesting capabilities are deployed with no knowledge of what to reliably expect, nor with any knowledge on how the energy will change with respect to factors in the surrounding network such as the transmitter density and the propagation channel.
\begin{figure}[t]
    \centering
    \includegraphics[width=1\linewidth]{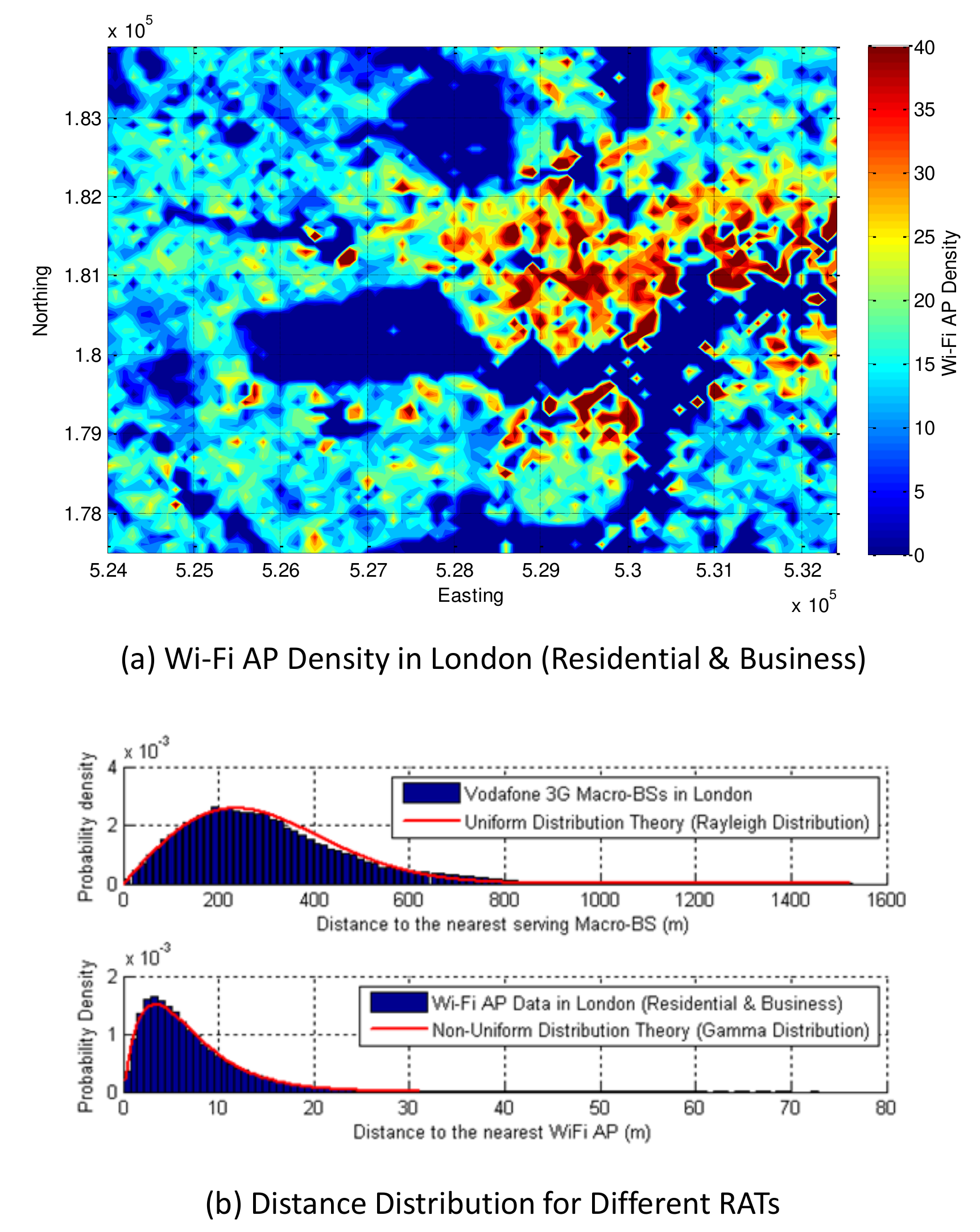}
    \caption{Case study of central London's 3G cellular and Wi-Fi networks: (a) density map of residential and business Wi-Fi APs, and (b) distance distributions from a random point to the nearest ($n=1$) macro-BS and Wi-Fi AP transmitter.  Data obtained from the United Kingdom (U.K.) Census (2011-15) and the The Office of Communications (Ofcom). }
    \label{Map}
\end{figure}

\subsection{Statistical Modeling: Distance Distributions}

Considering energy harvesting from a large number of fixed RF transmitters, it is possible to calculate the deterministic pathloss for each channel (i.e., ray-tracing) and predict the energy harvesting performance. However, several challenges exist, two of which are: i) \emph{computational complexity} grows linearly with the number of energy harvesting devices; and ii) \emph{imperfect knowledge of transmitters' locations} is a problem when private small-cells and mobile UEs are considered as energy harvesting sources. Hence, deterministically predicting the performance is challenging and statistical approximations maybe useful as bounds. Stochastic geometry studies random spatial patterns.  The underlying principle is that the locations of network transmitters are random in nature, but their spatial density and relative distances follow stable distributions. This can be used to create tractable statistical frameworks for analyzing the energy harvesting performance \cite{Guo13Wiley, Flint14}.

In order to estimate the energy received from a large number of RF transmitters, one needs to know the probability distribution of the distance between the ${n}$-th nearest RF transmitter and the energy harvesting device. To demonstrate this, let us consider for a moment that all the transmitters follow a certain spatial distribution with node density $\Lambda$. For example, there is a strong body of evidence that macro-BSs are distributed following a Poisson point process (PPP), i.e., each macro-BS is deployed independently to others. In the literature, the probability density function (pdf) of the distance $r$ between the energy harvesting device at an arbitrary location and the ${n}$-th nearest macro-BS is given by \cite{Guo13Wiley} $f_{\text{BS},{R}_{{n}}}({r};{n},\Lambda)$. Fig.~\ref{Map}(b)-top shows the ${n}=1$ case for PPP distributed macro-BSs, which follows a Rayleigh distribution.

As for small-cells (e.g., Wi-Fi APs and home femto-BSs), there is a lack of knowledge about the distance distribution due to the lack of large-scale empirical data. Two types of small-cells exist: (1) operator deployed, and (2) privately deployed. Existing research has largely focused on the former, whereby it has been proposed that the distribution of small-cells can follow Poisson cluster processes (PCPs). Using UK Census data as a proxy, we infer small-cell locations from household and business population and location data (UK Wi-Fi penetration is 89.8\%). The evidence gathered from a statistically significant amount of data indicates that the small-cell distance distribution is a non-uniform clustered spatial process, whereby the nearest distance distribution closely matches a Gamma distribution (see Fig. 3b). It remains an open area of research to explore the precise spatial process behind small-cells and the impact it has on crowd energy harvesting.

\subsection{Energy Harvesting Scaling Laws and Reliability}

In this subsection, we consider the aggregated RF power density (W/Hz) over a bandwidth of $B$ (Hz) and from an area with an average transmitter density of $\Lambda$.

\subsubsection{Upper-Bound (Full Spectrum Utilization) with Dual-Slope Pathloss}

\begin{figure}[ht!]
    \centering
    \includegraphics[width=1\linewidth]{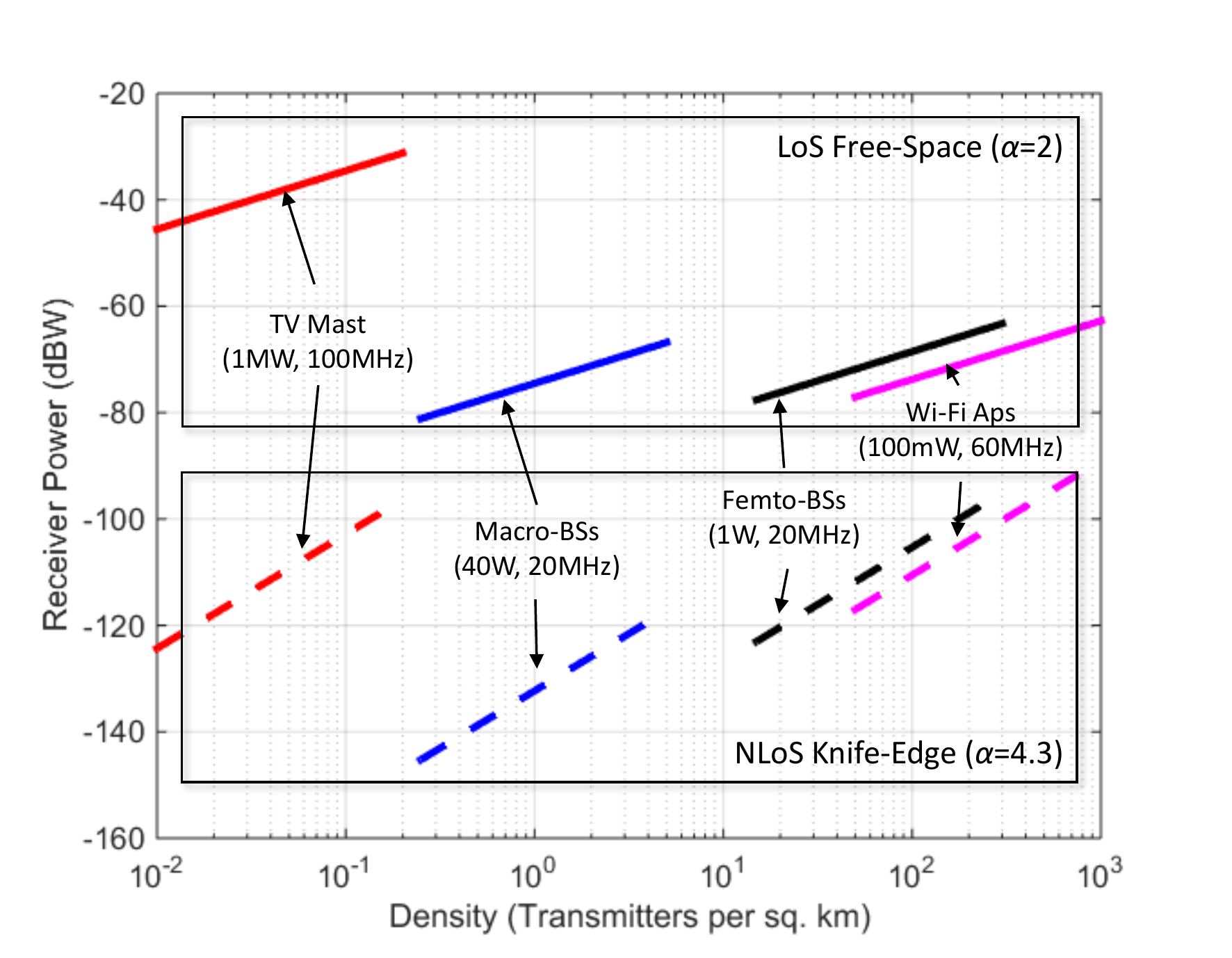}
    \caption{Upper-bound power harvested from different RATs as a function of transmitter density $\Lambda$ and pathloss scenarios.}
    \label{Bound}
\end{figure}

In the upper-bound analysis, we assume that: i) all transmitters are transmitting across the whole spectrum available, and ii) are emitting with the maximum allowable power spectrum density on all frequency bands. Leveraging the spatial distributions of RF transmitters found in the previous subsection, research in \cite{Guo13Wiley} found that the received power density $P_{\text{rx}}$ is linearly proportional to the bandwidth $B$ and the transmit power density $P_{\text{tx}}$, and has a complex monotonic relationship with the  distance-dependent pathloss exponent $\alpha$ and the node density $\Lambda$. Given that the energy from each transmitter will vary significantly depending on whether the propagation is largely Line-of-Sight (LoS, $\alpha=2$) or Non-LOS (NLoS, $\alpha>2$), one can consider a dual-slope approach (as shown in Fig.~\ref{Bound}), where two sets of pathloss exponents are considered \cite{Andrews15}: typically $\alpha=2$ for LoS free-space propagation, and $\alpha>4$ for NLoS urban propagation. As a result, the total power (averaged over distance) harvested from $K$ RF transmitters (each following its own spatial distribution) follows the following scaling rules \cite{Guo13Wiley}:
\begin{itemize}
  \item Linear with transmit power: $P_{\text{rx}} \propto P_{\text{tx}}$;
  \item Polynomial with transmitter density: $P_{\text{rx}} \propto (\Lambda)^{\frac{a}{2}}$;
\end{itemize}

As a case study, we consider the central London area (60 sq. km as shown in Fig.~\ref{Map}(a)) with network parameters for multiple radio-access technologies (RATs): i) cellular macro-BS downlink (20 MHz bandwidth, 40W, 0.3--5/km$^{2}$, real locations); ii) cellular femto-BS downlink (20 MHz, 1W, 15--200/km$^{2}$, PCP distributed); iii) Wi-Fi AP downlink (60 MHz, 100mW, 50--1000/km$^{2}$, proxy locations); and iv) TV broadcast (100 MHz, 1000kW, 0.01--0.2/km$^{2}$, real locations). The pathloss model considered is the WINNER model\footnote{The WINNER (Wireless World Initiative New Radio) is a statistical radio propagation model (2-6 GHz) for link and system level simulations of a variety of short range and wide area wireless communication systems.} with the appropriate shadow fading for different urban propagation environments.

Fig.~\ref{Bound} shows the upper-bound power harvested from different RATs as a function of transmitter density $\Lambda$ and different path-loss exponent values $a$. In particular, two exponent values are considered: (top) $a=2$ (free-space), and (bottom) $a=4.3$ (urban NLOS propagation in WINNER model). Table~\ref{Results} shows the peak harvested power (W) and power density (W/Hz) for different RATs (full spectrum usage). The results are obtained from extensive Monte-Carlo simulations in a manner similar to \cite{Guo13Wiley}. It can be seen that the greatest opportunity for power harvesting lies in the TV broadcast channels. Given that the network traffic of small-cells and TV is typically higher than that of macro-BSs, it is advisable to focus crowd energy harvesting in these bands for relaying.In Fig.~\ref{Bound}(top), the free-space (LoS) results match those found in existing field test observations.For example, it was found that 100~$\mu$W can be achieved at a 20km distance from a 150kW TV station~\cite{Ajmal14}. Looking ahead, we do not expect the node density for Macro-BSs and Wi-Fi hubs to change over the coming years, but the estimates are that the Femto-BS density will increase by at least 20 fold. Therefore, we expect a polynomial increase (exponent $a/2$, where the value of $a$ is typically 2-4) in the power available for harvesting. This potentially will lead to femto-BSs to act as an alternative or complementary source of crowd harvesting RF energy.

As for NLoS energy harvesting, as shown in Fig.~\ref{Bound}(bottom), the values are several orders of magnitude lower, due to the more significant distance dependent path loss, affecting the long range TV signals more than short-range Wi-Fi signals. Therefore, it is better to use Wi-Fi bands for NLoS energy harvesting, as opposed to TV. A key consideration from these results is that a relay performing RF harvesting would also require a band for information delivery, which should be the bands that present the least energy harvesting potential. Whilst the nearest node accounts for 89\% of the energy from crowd harvesting, in realistic networks, the nodes may not be transmitting at full-buffer continuously, and analysis that incorporates traffic patterns is necessary to understand the reliability advantage of crowd energy harvesting over nearest-node harvesting.
\begin{table}[t]
    \caption{Case study results with different RATs.}
    \begin{center}
        \begin{tabular}{|l|l|}
          \hline
          \textbf{RAT}                      & \textbf{Peak Power Density}                  \\
          Macro-BS Downlink                 & 11 fW/Hz                                     \\
          Femto-BS Downlink                 & 24 fW/Hz                                       \\
          Wi-Fi Downlink                    & 9 fW/Hz                                       \\
          TV Broadcast                      & 7550 fW/Hz                                     \\
          \hline
          \textbf{RAT}                      & \textbf{Peak Power (LoS)}                           \\
          Macro-BS Downlink                 & 0.21 $\mu$W                                     \\
          Femto-BS Downlink                 & 0.47 $\mu$W                                       \\
          Wi-Fi Downlink                    & 0.18 $\mu$W                                       \\
          TV Broadcast                      & 151 $\mu$W                                       \\
          \hline
        \end{tabular}
    \end{center}
    \label{Results}
\end{table}

\subsubsection{Realistic Traffic Load (Variable Spectrum Utilization)}

In order to estimate the realistic time-dependent RF energy, it is important to consider the spectrum utilization over time for each RF transmitter, which depends on the traffic load of each transmitter. Leveraging a well-known statistical model of 3G HSPA networks~\cite{Rupp12}, one can infer the spectrum utilization pattern at each BS as a ratio of the traffic and the peak capacity of the BS. Given that the $N$ BSs are independent and identically distributed in space and in spectrum utilization,the pdf of the power density (W/Hz) harvested from all $N$ RF transmitters is the linear combinations of random variables. This corresponds to the convolution of probability distributions if the traffic random variables are independent. Therefore, the $N$-fold \textit{continuous convolution} of the traffic load pdf, i.e., $f_{L_1} \ast f_{L_2} \ast f_{L_3} \ast ... \ast f_{L_N}$. These foundation statistical results and future research will provide useful guidelines to designing crowd harvesting powered relaying systems under variable spectrum utilization.

\section{Optimization for Crowd Harvesting}
\label{three}

We have so far reviewed SWIPT and non-SWIPT relaying, and in particular how non-SWIPT relaying can benefit from crowd RF energy harvesting across different RATs, which is attractive, especially in the TV bands (LoS) and Wi-Fi bands (NLoS).  However, what remains unclear is how a relay system, where the nodes are sufficiently apart (and hence have different energy harvesting potentials), can collaborate to achieve optimal relaying performance.  In this section we discuss node collaboration and transmission scheduling with QoS guarantees for relaying with crowd harvesting.

\subsection{Node Collaboration}
It has been revealed that the correlation distance of the traffic density is less than 80 meters in urban areas \cite{Dongheon}, indicating that the RF energy harvesting process may follow similar spatial correlation. Two nodes that are more than 100 meters apart tend to have almost independent energy harvesting processes, and thus node collaboration can be performed to exploit the independent relationship between energy profiles, e.g., to achieve energy harvesting diversity gains.

First we illustrate the benefit of node collaboration via combining the SWIPT and ambient energy harvesting, in order to compensate for the possible energy shortage at either source or relay as shown in Fig.~\ref{fig-YC1}. Assuming that the source can harvest more ambient RF energy than the relay node, one can introduce new TS parameters $\alpha_1$ and $\alpha_2$ ($0 \le \alpha_1$, $0 \le \alpha_2$, $\alpha_1 + \alpha_2 \le 1$), whereas the energy harvesting phase can be split further into two parts, with $\alpha_1 T$  seconds and $\alpha_2 T$ seconds, for: i) source-to-relay energy delivery, and ii) ambient RF energy harvesting at relay, respectively.  Note that the source can make use of the time when the relay forwards the message, to harvest additional ambient RF energy. For PS, a similar protocol can be adopted, with power-splitting factors $\rho_1$ and $\rho_2$ ($0 \le \rho_1$, $0 \le \rho_2$, $\rho_1 + \rho_2 \le 1$) for source energy delivery and ambient RF energy harvesting, respectively. While this may require another energy harvesting component, a mechanism of hybrid TS and PS can be a promising solution without additional hardware requirements.

\begin{figure}[t]
    \centering
    \includegraphics[width=1\linewidth]{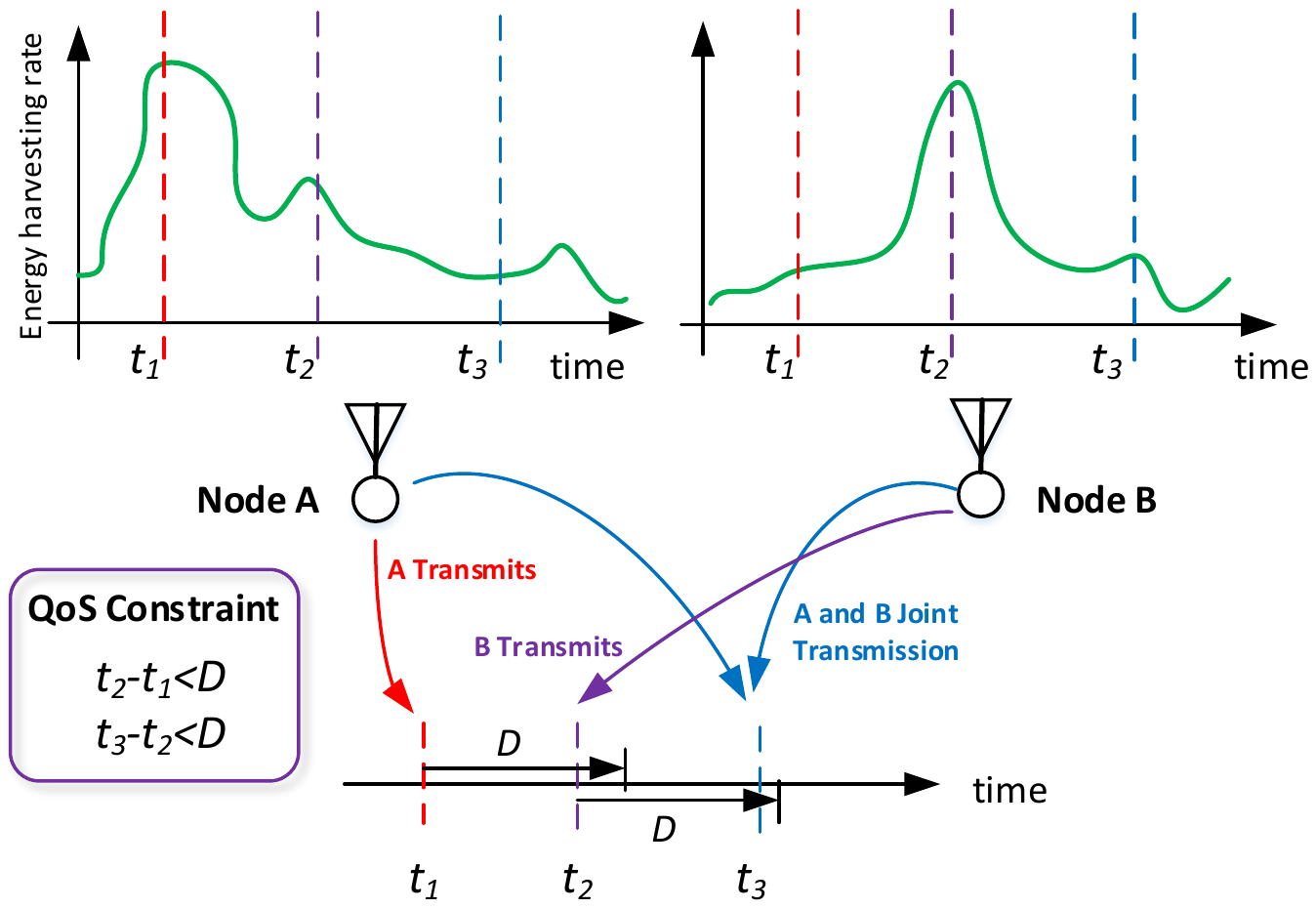}
    \caption{Transmission scheduling and node collaboration under independent energy arrival processes and inter-delivery time requirement.}
    \label{Col}
\end{figure}

Furthermore, for the scenario where the nodes have no SWIPT structure or have some common information to the same destination, for example, the multiple relays in the second hop of a relaying transmission, or multiple sensors that sense the same target and need to deliver the sensing results to the sink. In this case, collaborative transmission can be used to address the uneven energy arrival rates, for instance the energy arrival process illustrated in Fig.~\ref{Col}, for node A and B. As a simple example, a transmission frame can be divided into two subframes. In the first subframe, only one of the nodes can be scheduled to transmit in the conventional way. In the second subframe, multiple nodes can perform simultaneous joint transmission (JT) to the destination with distributed beamforming. To this end, the frame division portion $\xi$ ($0 \le \xi \le 1$) and the node scheduling should be jointly optimised, taking into account the ESI of all nodes.

For both cases, to get the optimal system parameters $\alpha_i$, $\rho_i$ and $\xi$, practical online algorithms should be designed based on the prediction of the mobile traffic that generates the crowd EH source. One can model the mobile traffic variation with Markovian model, of which the transition probabilities can be trained based on real data as presented in Section \ref{two}, and then MDP policy iteration will provide the optimal transmission scheduling and system parameters. To reduce the complexity of MDP, one can do conventional optimization on a per-frame basis, with the energy arrivals and channel conditions of several future frames as known, given the traffic prediction precision is high.

\subsection{Delay QoS Guarantee}

As a type of \emph{delay-sensitive} traffic, the transmission of the sensing data is subject to a temporal regularity constraint, for example the inter-delivery time requirement \cite{Xueying}, as a large gap between updates can cause system instability. To guarantee such delay QoS requirement, the transmission scheduling should utilize the \emph{diversity} from the different energy harvesting processes of nodes. For instance in Fig.\ref{Col}, there are three consecutive delivery instants, $t_1$, $t_2$ and $t_3$, and it is required that the inter-delivery time should be less than $D$. At time $t_1$, node A has a high energy harvesting rate and can afford to deliver the sensing data to the sink, and likewise at time $t_2$, node B has enough energy to deliver the data. However at time $t_3$, both nodes have low energy arrival rates and the inter-delivery time constraint $D$ is about to break, thus they use collaborative JT to ensure reliable transmission to the sink.

Some delay-insensitive applications require that important events are sensed or sampled with finer time granularity, but can be delivered less timely. In this case, the sensor node needs to act proactively according to the energy stored in its battery and the prediction of future energy arrivals. If the battery is about to overflow, the node can send the sensing results in a batch to the destination/sink, while maintaining enough energy for possible sensing efforts when important events happen. By optimizing the energy storage in the battery and the scheduling of transmission/sensing activities, the node can equivalently \emph{match} the energy arrivals and the occurrence of sensing events over time.

\section{Open Challenges}
\label{four}

WPR is a relatively new technique. It has strong potential for the emerging M2M communications in IoT systems, but a number of open challenges exist. Accurate \emph{energy modeling} is important. Most existing work merely considers the energy used for transmission. However, the energy consumption for receiving is comparable with that of transmission and cannot be ignored \cite{Sheng}. Therefore, scheduling decisions should take into account the energy consumption of receiving as well, and a relay node should switch among receiving, forwarding, and keeping silent modes to save energy, based on its ESI, CSI of the multiple hops, and the delay constraints. In terms of optimization, an area of interest that has not been mentioned in this paper is \emph{relay selection}, which provides a trade-off between complexity and performance. However, the relay selection criterion for information transmission may be different from that for energy transfer. For example, previous research has shown that the relay node offering the largest end-to-end SNR may not be the one with the largest received power \cite{Chen}. In SWIPT WPR, it is not clear which selection criterion should be used to achieve the best performance. In practice, the distances between nodes will significantly affect the system performance, as existing energy harvesters on work efficiently over short distances. Therefore, a practical challenge is to decide whether it is better to use multi-hop with shorter distances.

%In summary, we have reviewed recent progress in simultaneous energy and information transfer for relay networks. Two types of RF powered relay systems are considered: those harvesting energy from the source node's transmissions, and those crowd-harvesting RF transmissions. Both can greatly reduce the use of battery power and increase the availability and reliability for relaying. Through a case study, we show that it is better to harvest energy from TV bands in LoS channels, but better to harvest energy from ambient Wi-Fi signals in NLoS conditions. Furthermore, we have discussed the optimization of transmissions in crowd harvesting, especially with the use of node collaboration for delay QoS guarantee.

\bibliographystyle{IEEEtran}
\bibliography{IEEEabrv,Bib}
%%%%%%%%%%%%%%%%%%%%%%%%%%%%%%%%%%%%%%%%%%%%%%%%%%%%%%%%%%%%%%%%%%%%%%%%%%%%%%%
%%%%%%%%%%%%%%%%%%%%%%%%%%%%%%%%%%%%%%%%%%%%%%%%%%%%%%%%%%%%%%%%%%%%%%%%%%%%%%%

\end{document}